\documentclass[final,conference,twocolumn,letterpaper]{IEEEtran}

\IEEEoverridecommandlockouts              
\usepackage{fancyhdr}
\usepackage{graphicx}
\usepackage[utf8]{inputenc}

\graphicspath{{figures/}}
\usepackage{url}
\usepackage{wrapfig}
\usepackage{units}
\usepackage{caption}
\usepackage[printonlyused]{acronym}

\usepackage{calc}
\usepackage{graphicx}
\usepackage[lined,boxed, vlined,ruled,linesnumbered]{algorithm2e}
\usepackage{amsmath,amssymb,amsthm,amsfonts,amsbsy}

\usepackage{algpseudocode}
\usepackage{color}
\usepackage[table]{xcolor} % DVI only
\usepackage[caption=false]{subfig}
\usepackage{dblfloatfix}
\usepackage{nicefrac}
\usepackage{tcolorbox}
\usepackage{txfonts}

\usepackage{multirow}

\newcolumntype{C}[1]{>{\centering\arraybackslash}p{#1}}

\usepackage{booktabs}

\definecolor{BgGray}{gray}{0.7}%
\definecolor{BgGray2}{gray}{0.96}%
\definecolor{RowColorOdd}{named}{BgGray2}%
\definecolor{RowColorEven}{named}{white}%
\definecolor{comments}{gray}{.5}
\definecolor{Gray}{gray}{0.85}

\providecommand{\keywords}[1]{\textbf{\textit{Index terms---}} #1}

\usepackage{pifont}% http://ctan.org/pkg/pifont
\usepackage[multiuser,nomargin,marginclue,footnote]{fixme}
\fxusetheme{color}
\fxsetup{targetlayout=colorcb}
\fxsetup{status=draft}

\begin{document}

\title{Practical MIMO for Visible Light Communication}

\author{
\IEEEauthorblockN{Piotr Gawłowicz, Elnaz Alizadeh Jarchlo, Anatolij Zubow}
\IEEEauthorblockA{\{gawlowicz, jarchlo, zubow,\}@tu-berlin.de}
Technische Universität Berlin, Germany}

\maketitle

%%%
\begin{abstract}
Visible Light Communication (VLC) is seen as a complementary wireless technology to Radio Frequency (RF).
However, VLC is very sensitive to the signal blockage and suffers from shadowing due to the high directionality of the optical channel. 
Hence there is a big interest in researching novel approaches for VLC like usage of multiple antenna techniques providing spatial diversity which can be exploited as a way to combat signal blockage and fading.
We present a complete and low-cost MIMO-VLC transceiver system consisting of COTS components.
In particular, we show that COTS 802.11n (WiFi) devices can be used so that the physical and data link layers of RF WiFi are reused for VLC.
In addition, this allows us to directly utilize the multiple antenna (spatial) techniques available in 802.11n.
Results from our measurement study show that such techniques are highly effective at improving the robustness of VLC links in the presence of obstacles and node mobility.
Moreover, we show that multiple antennas can also be used to increase the data rate of VLC by means of spatial multiplexing.
\end{abstract}

\keywords{OWC, VLC, WiFi, MIMO, spatial diversity, spatial multiplexing, COTS}

%
%%%
\section{Introduction}
VLC is a short-range wireless communication technology that has gained substantial attention from both industry and the research community in both indoor and outdoor network environments.
Besides illumination, VLC uses light-emitting diodes (LEDs) in order to transmit the signal with a high rate of modulation that is not detectable by the human eyes.
Although VLC has several advantages~\cite{jovicic2013visible} such as its high security, a high degree of spatial reuse and increased capacity thanks to its large spectrum (i.e. high bandwidth) which implies good spatial resolution and energy and cost-efficiency, it faces several challenges as well.
For instance, VLC suffers from connectivity disruption due to shadowing, blockage, mobility and, the external light~\cite{ghassemlooy2019optical}. 

IEEE 802.11 has an essential characteristic specification, i.e. a single medium access control (MAC) sub-layer common to many physical (PHY) layers. 
This feature provides an opportunity for easier interoperability among the physical layers that are already or will be available in the future.
We believe that VLC is the desired candidate as a complementary technology to RF WiFi.
Moreover, 802.11 is already very advanced technology.
Since the introduction of the 802.11n amendment, WiFi is based on a multiple antenna physical layer that supports multiple input multiple output (MIMO) techniques.
These techniques have generated much excitement due to predictions of large capacity gains.
We believe that MIMO will also play a significant role in VLC as techniques like spatial diversity, i.e. combining the signal from multiple antennas, are very effective at improving the robustness of VLC links in presence of obstacles and node mobility.
Moreover, spatial multiplexing can increase aggregate throughput of VLC systems by sending different signals simultaneously using multiple uncorrelated optical channels.

In this paper, we investigate how well the 802.11n physical layer works over an optical (VLC) channel.
First, we present a complete and low-cost MIMO-VLC transceiver system consisting of COTS components.
A low-cost solution is of great importance in order to be able to build large scale MIMO-VLC systems, i.e. in terms of the number of nodes and antennas (or VLC front-ends).
Thereafter, the results of a measurement study from our small indoor testbed are presented.
Our results reveal that the spatial diversity technique from 802.11n is very efficient in improving the robustness of VLC in case of signal blockage due to obstacles.
Moreover, MIMO spatial multiplexing is applicable even if the VLC nodes (TX/RX) are just slightly separated in space.
Finally, as capturing channel state information is possible with WiFi NICs such information can be used to study channel characteristics of VLC.

\smallskip

\noindent\textbf{Contributions:}
We present MIMO-WoV (MIMO-\textbf{W}iFi-\textbf{o}ver-\textbf{V}LC), a simple and inexpensive COTS-based evaluation platform for prototyping novel MIMO-VLC solutions.
Specifically, we exploit the vast set of capabilities, i.e. MIMO spatial diversity and multiplexing, already implemented in modern RF 802.11 chipsets over optical links.
Therefore, a researcher can focus entirely on the design of novel VLC front-end (FE) solutions.
We provide proof of concept prototype implementation and perform an evaluation in a small testbed.
As the proposed hardware is cheap even large-scale VLC testbeds can be built to analyze the performance of VLC on both link as well as system-level.
\section{Background}

As background, we briefly discuss the multiple antenna (MIMO) techniques of spatial diversity and multiplexing.
Then, we discuss the feasibility and support of those techniques in the WiFi standard and VLC technology.

\subsection{Spatial Diversity and Multiplexing}
\textbf{Spatial Multiplexing (SM)} is a transmission technique in Multiple-Input and Multiple-Output (MIMO) wireless communication to transmit independent and separately encoded data signals called \textit{streams}. 
Therefore, the space dimension is reused, or multiplexed, more than one time.
If the transmitter is equipped with $N_t$ antennas and the receiver has $N_r$ antennas, the maximum spatial multiplexing order (the number of streams) equals $N_s = \min(N_t, N_r)$ if a linear receiver is used~\cite{tse2005fundamentals}.
This means that $N_s$ streams can be transmitted in parallel, ideally leading to an $N_s$ increase of the spectral efficiency.
In a practical system, the multiplexing gain is often limited by spatial correlation, which means that some of the parallel streams may have very weak channel gains.
Direct-mapping~\cite{halperin2009two} is the simplest MIMO technique where each antenna transmits its own data stream. %, which is used in 802.11n.
In a rich scattering RF environment where $m$ transmit streams are received by $n$ antennas, each receive antenna will measure an independent linear combination of the $m$ signals.
This is decodable when $n \geq m$ so that there are more measurements ($n$) than unknowns ($m$).
Extra measurements ($n > m$) add diversity gain.
A MIMO receiver uses zero-forcing techniques to solve the MIMO linear system.
Note, that usually (e.g. in WiFi) all streams use the same rate (MCS) and same Tx power.

Next, we have two types of \textbf{Spatial Diversity (SD)}.
Here we can distinguish between receive diversity, using multiple receive antennas (single input multiple output or SIMO channels), and transmit diversity, using multiple transmit antennas (multiple input single output or MISO channels).
SIMO techniques like Maximal Ratio Combining (MRC) are used to harness the useful power from all antennas by adding the signals from different antennas in a coherent manner~\cite{halperin2009two}. 
Therefore, the receiver needs an estimate of the channel effect (between each pair of TX and RX antennas), which it obtains by using training fields in the preamble. MRC reverts the effect of the channel, i.e. it delays signals from different antennas so that they have the same phase, weights them proportionally to their SNR, and adds them. 
In contrast, the Selection Combining (SC) technique simply selects a signal with the highest Rx power.
Finally, spatial diversity can be also obtained on the transmit-side.
Here the sending node can either select the best antenna to transmit or it can use beamforming techniques to make sure that different signal copies combine coherently at the receiver side~\cite{halperin2009two}.
However, transmit diversity requires channel knowledge on the transmitter side, hence typically relying on receiver feedback.

\subsection{MIMO in WiFi}
MIMO is an integral part of WiFi since 2009 when the 802.11n standard amendment was published.
Most 802.11n NICs support both spatial receive diversity (via MRC) and up to $4\times4$ multiplexing (via direct-mapped MIMO).
Transmit diversity (i.e. beamforming) is an optional feature in 802.11n, however, it is mandatory in newer generations. MIMO dimensions were extended by the 802.11ac amendment to $8\times8$.
Staring from the 802.11n version, the physical layer is based on OFDM, while 802.11ax introduces OFDMA.

\subsection{MIMO in VLC}

The usage of MIMO techniques in VLC is highly desired especially in manufacturing environments as well as in vehicular communication, where the spatial diversity techniques are essential in order to achieve required low latency and bit error rate~\cite{berenguer2017optical}. 
Moreover, MIMO-VLC is also expected to improve user experience in home environments, where its development and expansion might be even easier as most of the luminaires typically contain multiple LEDs that can be used as multiple transmitters~\cite{AlAhmadi2018MultiUserVL}. 
However, to the best of our knowledge, there is currently no commercial VLC device that supports MIMO.

%
%%%
\section{System Design} \label{sec:arch}

This section gives an overview of our proposed MIMO-\textbf{W}iFi-\textbf{o}ver-\textbf{V}LC (MIMO-WoV) architecture. 
First, we briefly introduce our previous platform, namely WoV, that serves as a basis for MIMO-WoV.
Then, we describe the introduced modifications that enabled the usage of MIMO capabilities.

\begin{figure}[ht]
    \centering
    \vspace{0pt}
    \includegraphics[width=0.8\linewidth]{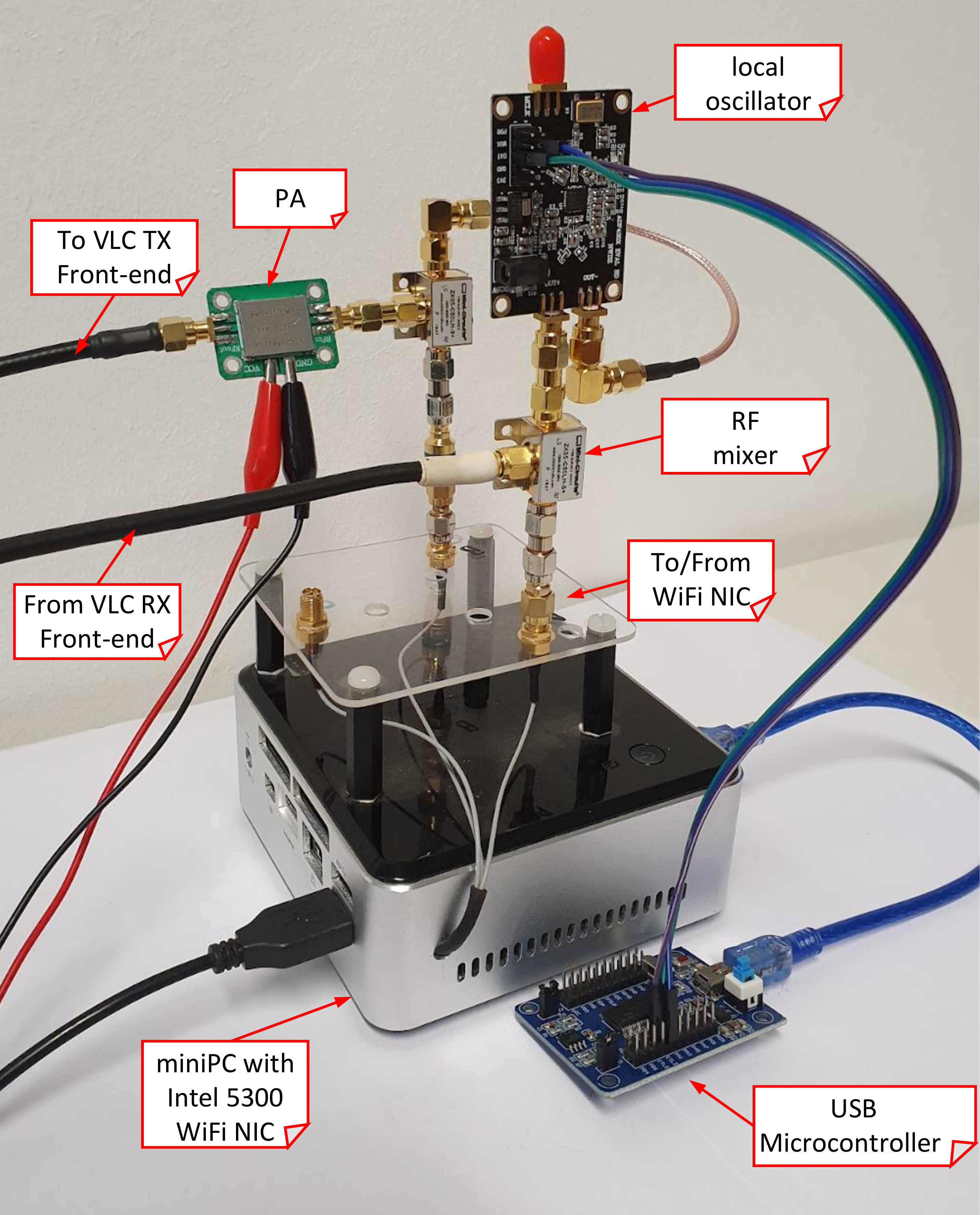}
    %\vspace{-5pt}
    \caption{WiFi-over-VLC (WoV) node used as a basis.}
    \label{fig:tx_prototype}
    \vspace{-5pt}
\end{figure}
\subsection{Overview of WoV}
In~\cite{gawowicz2020wov} we presented the WiFi-over-VLC (WoV) architecture and showed that the physical and data link layers of 802.11 WiFi designed for RF can be reused in VLC.
Specifically, we demonstrated that the RF signals emitted by COTS 802.11 WiFi NICs can be used for direct transmission over optical wireless links.
A WoV node supporting bi-directional SISO communication is presented in Fig.~\ref{fig:tx_prototype}. 
It consists of a host PC (i.e. Intel NUC mini PC), equipped with a single COTS WiFi network interface cards (NIC) (i.e. Intel WiFi Wireless Link 5300), that generates and consumes 802.11-compliant waveforms.
The base-band signals from unmodified WiFi NICs are down-converted to match the input specification of VLC front-end modulates. 
The signal received from VLC RX front-end is up-converted so it can be directly injected into the WiFi NIC.
The up-/down-conversion is performed in the RF setup connected to two antenna ports of the NIC.
The setup consists of inexpensive components like a variable local oscillator (VLO) and two RF mixers (i.e. one for each TX and RX path).
Note that both conversions are required as COTS WiFi chipsets integrate baseband processing unit and radio transceiver in a single system-on-chip (SoC) and expose only RF signal in 2.4\,GHz or 5\,GHz band.
The RF ports of mixers is connected to VLC front-ends -- Fig.~\ref{fig:vlc_frontends}. In addition, the signal on the TX path is amplified to match the power specification of the TX front-end.

The proposed integration is \textit{transparent} to the higher layer of the protocol stack as well as to the WiFI NIC itself. Therefore, no modifications to the software side were required, i.e. we use the standard Linux OS and WiFi drivers.

\subsection{Extension towards MIMO}\label{section:hardware}
\begin{figure}[ht]
    \centering
    \includegraphics[width=\linewidth]{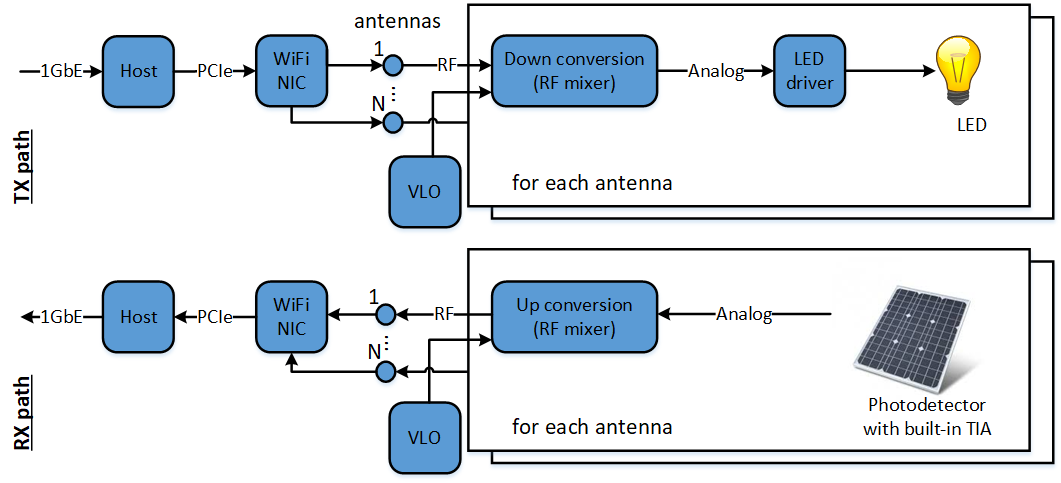}
    \vspace{-5pt}
    \caption{Architecture of MIMO-WoV.}
    \label{fig:wifi_over_vlc_arch}
    \vspace{-10pt}
\end{figure}

Although the basic WoV node utilizes two TX/RX transmission chains, it supports only a SISO bi-directional link, i.e. one chain is used only in TX, and one only in RX mode.
In order to enable the use and evaluate the performance of underlying WiFi MIMO capabilities, we modify the basic WoV architecture in the following way.
We use two WoV nodes, i.e. to the first one, we connect only TX VLC front-ends, while to the second one only RX VLC front-ends.
Moreover, we modify the configuration of the NICs and force all transmission chains to operate in only TX or RX mode, respectively. 
Fig.~\ref{fig:wifi_over_vlc_arch} shows the schematic diagram of the proposed changes. 
This way, we can immediately enable and evaluate advanced signal processing techniques (i.n. MIMO) implemented in the COTS WiFi NIC over VLC links, but at the cost of bi-directionality, i.e. the enabled MIMO link is uni-directional. We discuss this issue in detail in \S \ref{sec:discussion}.
Note that with the current prototype, we can enable bi-directionality by the implementation of an asymmetric system, where one MIMO-VLC node can use two TX front-ends and one RX front-end (i.e. three transmission chains in total), while the second node uses two RX front-ends and one TX front-end.

%
%%%
\section{Evaluation}
\begin{figure}[ht]
    \vspace{-5pt}
    \small
    \centering
    \includegraphics[width=0.8\linewidth]{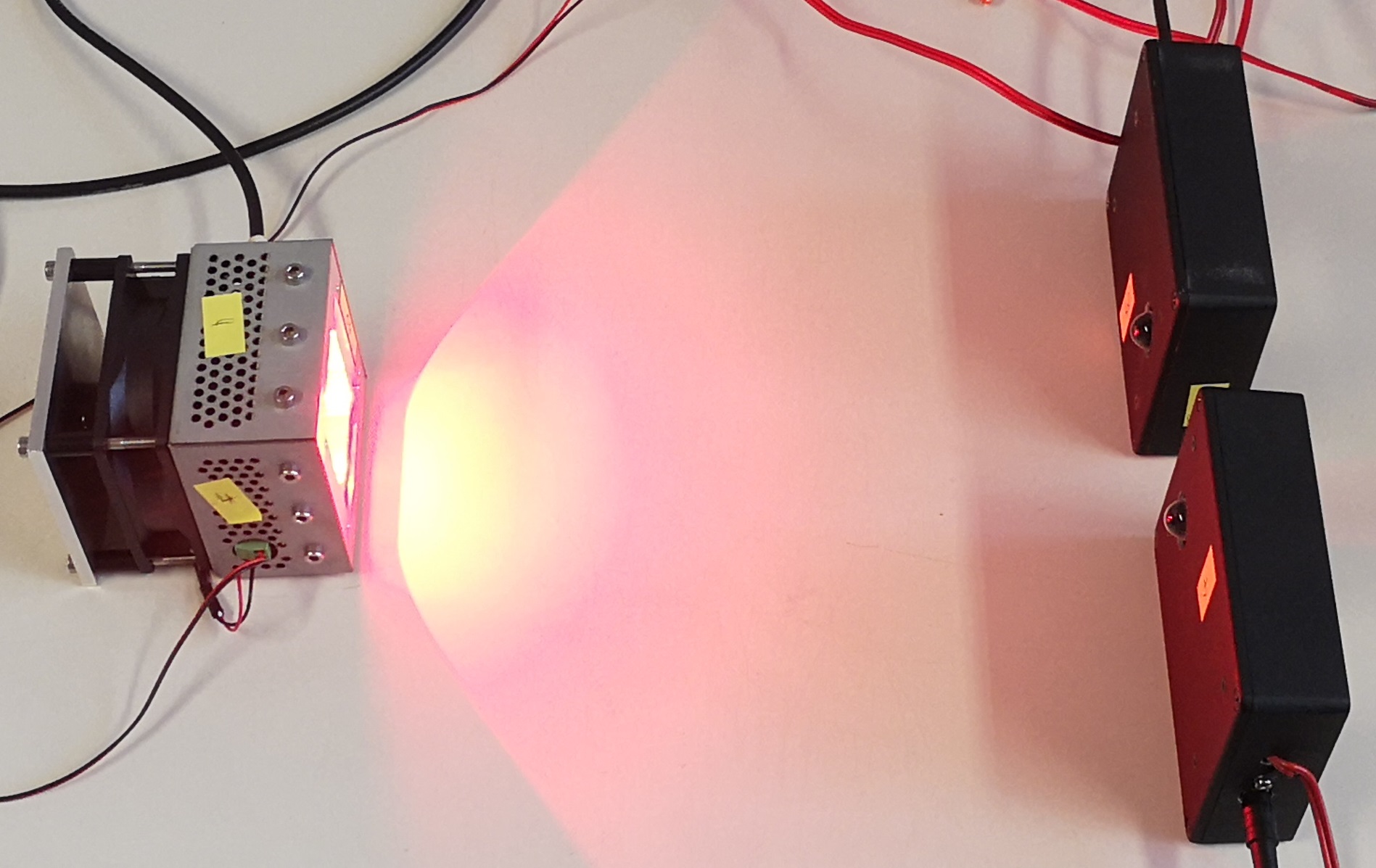}
    %\vspace{-2pt}
    \caption{VLC front-ends in a SIMO setup: single transmitter (left) and two receivers (right).}
    \label{fig:vlc_frontends}
    \vspace{-10pt}
\end{figure}

In this section, we present results from experiments using our MIMO-VLC prototype.
First, as a baseline, we present the performance of a single unidirectional SISO-VLC link.
Second, we investigate the efficacy of spatial diversity on the receiver side (SIMO).
Here we show that the SIMO technique available in RF 802.11n COTS WiFi can be used for VLC as well for making it robust against signal blockage due to obstacles and to improve mobility by providing soft-handover.
Third, we show that spatial multiplexing is possible, hence increasing the data rate.
All experiments are performed in a small testbed that consists of one TX node and one RX node with a varying number of VLC front-ends.
An example experiment setup used for evaluation of spatial diversity is shown in Fig.~\ref{fig:vlc_frontends}, where we see one TX VLC front-end and two RX VLC front-ends (SIMO).

\subsection{SISO Performance}
The purpose of this experiment it two-fold.
First, we show that we can successfully transmit WiFi signals over optical wireless links using our WoV testbed.
Second, we evaluate the performance of our setup in SISO link configuration, i.e. we use only single VLC transmitter and receiver front-ends each connected to the first antenna port of the two WiFi NICs.
During the experiment, we send 802.11n frames using different modulation and coding schemes (MCS) and measure the frame success rate (FSR). 
Moreover, we place the VLC front-ends at different distances to influence the received power strength (RSSI). 
For each measurement point, we transmitted 1000 frames of the size of 1000\,Bytes.

\begin{figure}[ht!]
  \centering
  \hfill
  \begin{minipage}[ht!]{0.20\linewidth}
    \includegraphics[width=1.0\linewidth]{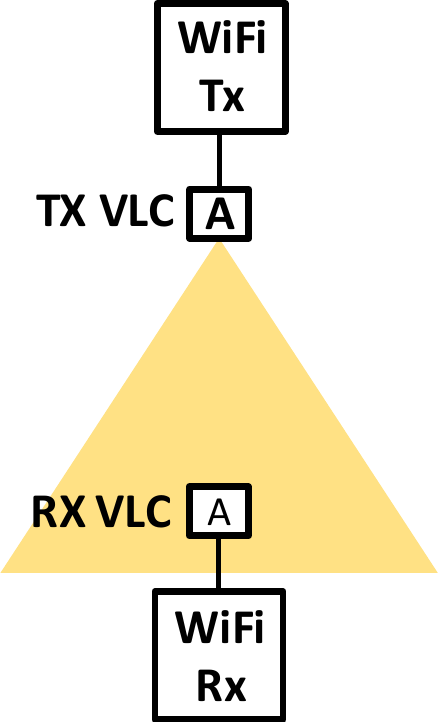}
    %\caption{Experiment Setup.}
  \end{minipage}\hfill
  \begin{minipage}[ht!]{0.75\linewidth}
    \includegraphics[width=1.0\linewidth]{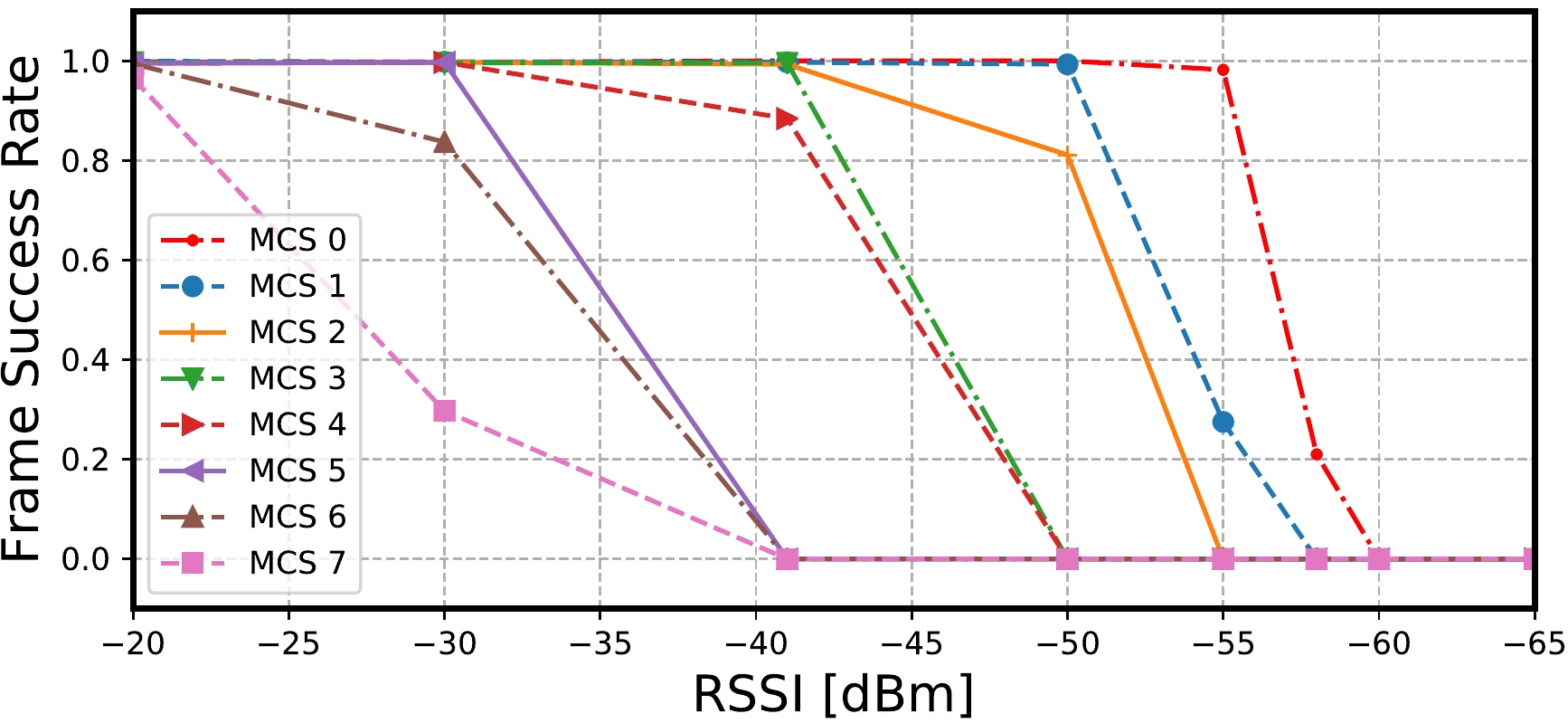}
  \end{minipage}\hfill
  \caption{SISO performance: (left) experiment setup, (right) Frame Success Rate vs. RSSI}
  \label{fig:mimo_vlc_exp_siso}
  \label{fig:fsr_sisio_20m}
  \vspace{-5pt}
\end{figure}

\noindent\textbf{Results: }
In Fig.~\ref{fig:fsr_sisio_20m} we show the relationship between FSR and RSSI for all MCSs between 0 and 7 as defined in 802.11n standard for single stream communication.
We see that the frames send using the MCS 0 (i.e. BPSK 1/2) can be received without any errors when the RSSI is higher then -55\,dBm. From this result and the known SNR thresholds for different 802.11n MCS~\cite{halperin2011predictable}, we can conclude that the noise floor of the WoV setup is around -60\,dBm.
Moreover, as expected, the same level of FSR for the MCS 7 requires the signal's SNR to be around 30\,dB higher.

\subsection{SIMO Performance (Spatial Diversity)}
The objective of the following experiment is to analyze two things.
First, we want to show that the receive spatial diversity technique from 802.11n is operational when using the VLC medium.
In particular, we would like to quantify the gain from the MRC technique.
Second, we want to study the suitability of that type of diversity as a way to combat signal blockage due to obstacles.
Therefore, the experiment is composed of a transmitter and a receiver node.
The former had a single antenna port connected to the VLC front-end transmitter whereas at the receiver side two antenna ports are used and each connected to a different VLC front-end receiver (see Fig.~\ref{fig:vlc_frontends}).
During the experiment, we randomly cover one of the VLC RX front-ends in order to block the communication link. 
Again, a one-way communication was set-up and unicast frames with MCS=0 (802.11n, BPSK) were injected.
A video demonstrating this experiment setup can be found here\footnote{\url{https://www.youtube.com/watch?v=cIuXZWFWaWY}}.

\begin{figure}[ht!]
  \centering
  \hfill
  \begin{minipage}[ht!]{0.25\linewidth}
    \includegraphics[width=1.0\linewidth]{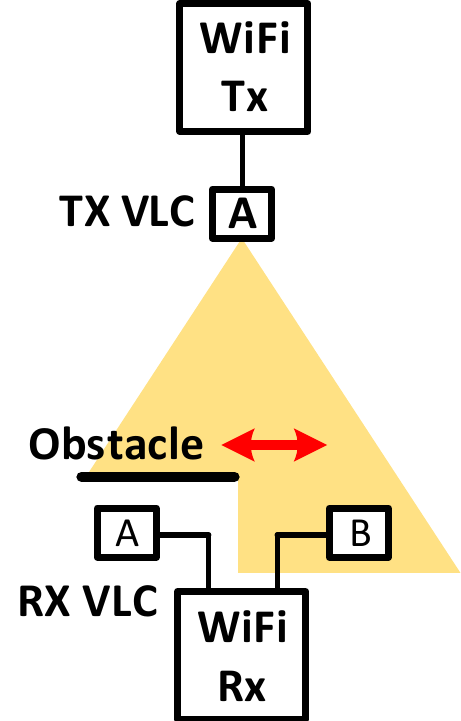}
  \end{minipage}\hfill
  \begin{minipage}[ht!]{0.75\linewidth}
    \includegraphics[width=1.0\linewidth]{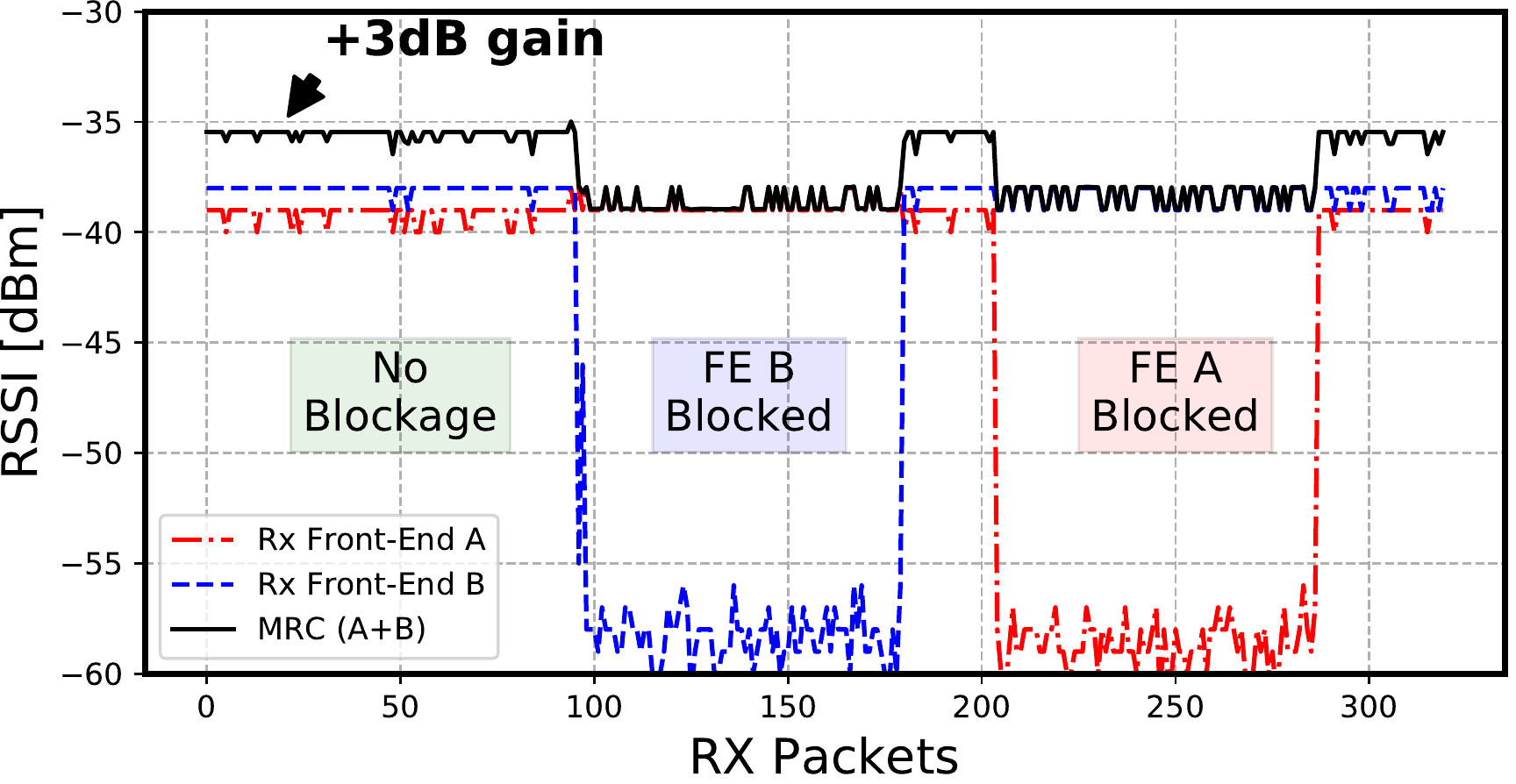}
  \end{minipage}\hfill
  \caption{Receive diversity (SIMO): (left) experiment setup, (right) RSSI for each received packet.}
  \label{fig:mimo_vlc_exp_blockage}
  \label{fig:rx_diversity}
  \vspace{-5pt}
\end{figure}

\noindent\textbf{Results: }
Fig.~\ref{fig:rx_diversity} shows the signal power level for all received frames on both paths $T_A-R_A$ and $T_A-R_B$, separately as well as the one after combining the signals from both paths using MRC (spatial diversity) for 350 transmitted packets.
During the transmission of the first 100 packets, there was no signal blockage, i.e. both VLC links had clear LOS conditions.
Here we can see that MRC gives us a gain of around 3\,dB.
During the reception of packets 100 to 180, the path $T_A-R_B$ was fully blocked.
Thanks to MRC the packets could be received on the second non-blocked path $T_A-R_A$.
Due to the full blockage of the RX front-end B, there was no power gain from MRC over path $T_A-R_A$ .
Finally, for the packets 200 to 280 the path $T_A-R_A$  was blocked whereas the RX front-end B had clear LOS towards TX front-end.
We can clearly see the advantage of spatial diversity using MRC, i.e. as long as at least one VLC path is available, the packet flow is not interrupted.

\begin{figure}[ht!]
  \centering
  \hfill
  \begin{minipage}[ht!]{0.25\linewidth}
    \includegraphics[width=1.0\linewidth]{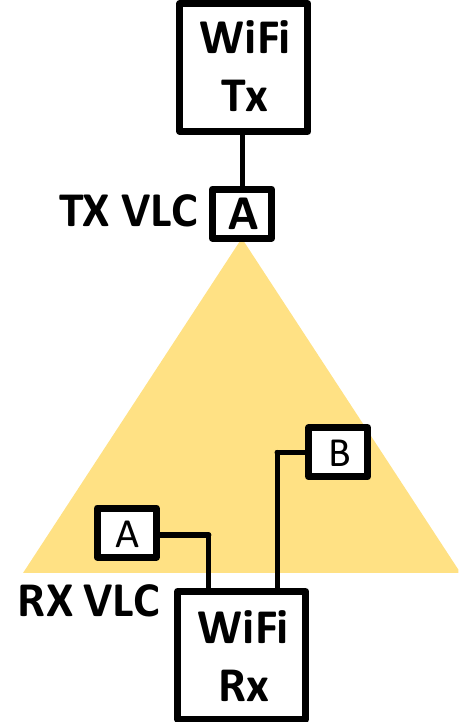}
    %\caption{Experiment Setup.}
  \end{minipage}\hfill
  \begin{minipage}[ht!]{0.7\linewidth}
    \includegraphics[width=1.0\linewidth]{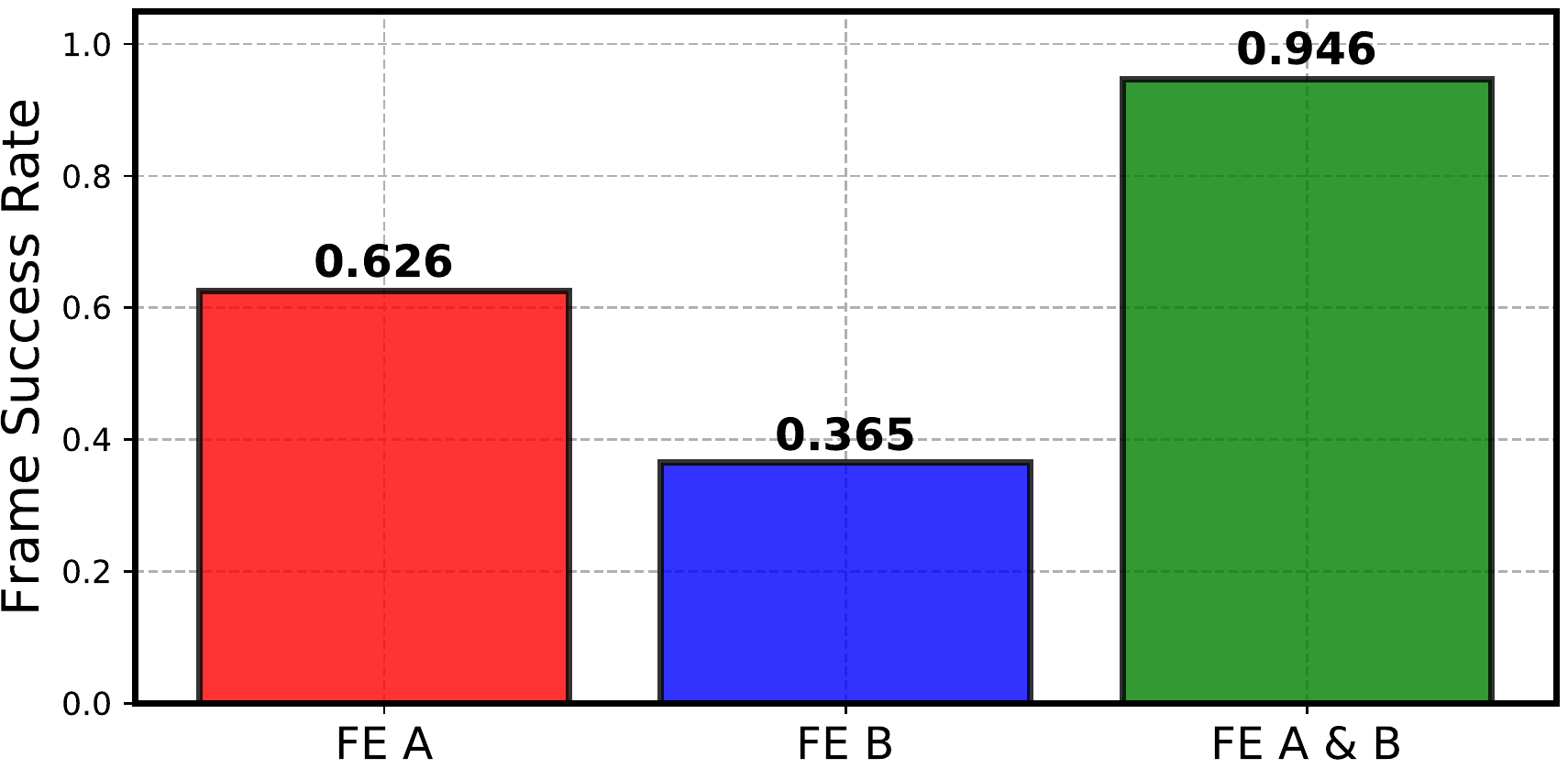}
  \end{minipage}\hfill
  \caption{Frame Success Rate at particular location for: \textit{i+ii)} SISO using front-end A and B separately, \textit{iii)} SIMO with MRC combining signals of both front-ends.}
  \label{fig:mimo_vlc_exp_mrc}
  \label{fig:rx_fer}
  \vspace{-5pt}
\end{figure}

Fig.~\ref{fig:rx_fer} shows the frame success rate (FSR) for arbitrary selected bad positions for the two RX VLC front-ends.
Although both have still clear line-of-sight (LOS) they suffer from low FSR due to weak signals.
The FSR for paths $T_A-R_A$ and $T_A-R_B$ is 0.626 and 0.365 respectively.
However, after combining using MRC the FSR was dramatically increased to 0.946.
This example shows that the MRC gain can make a difference when operating in the noise-limited region.

\subsection{Distributed Antennas for Soft Handover}
\begin{figure}[ht]
  \centering
  \hfill
  \begin{minipage}[ht!]{0.25\linewidth}
    \includegraphics[width=1.0\linewidth]{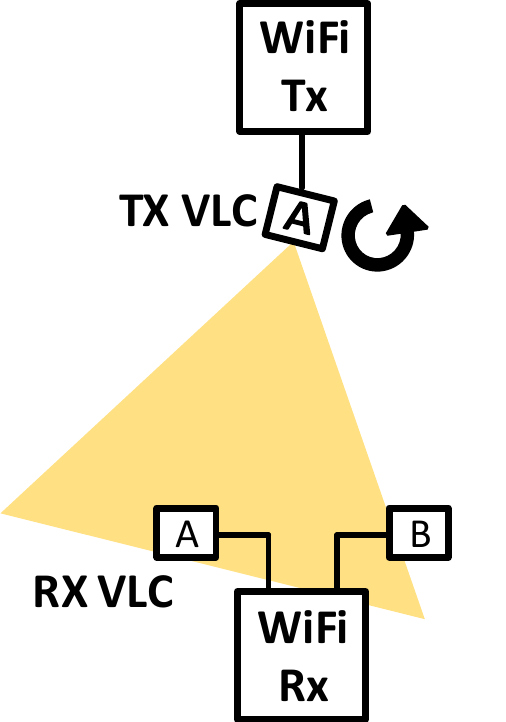}
    %\caption{Experiment Setup.}
    \label{fig:mimo_vlc_exp_rx_div}
  \end{minipage}\hfill
  \begin{minipage}[ht!]{0.75\linewidth}
    \includegraphics[width=1.0\linewidth]{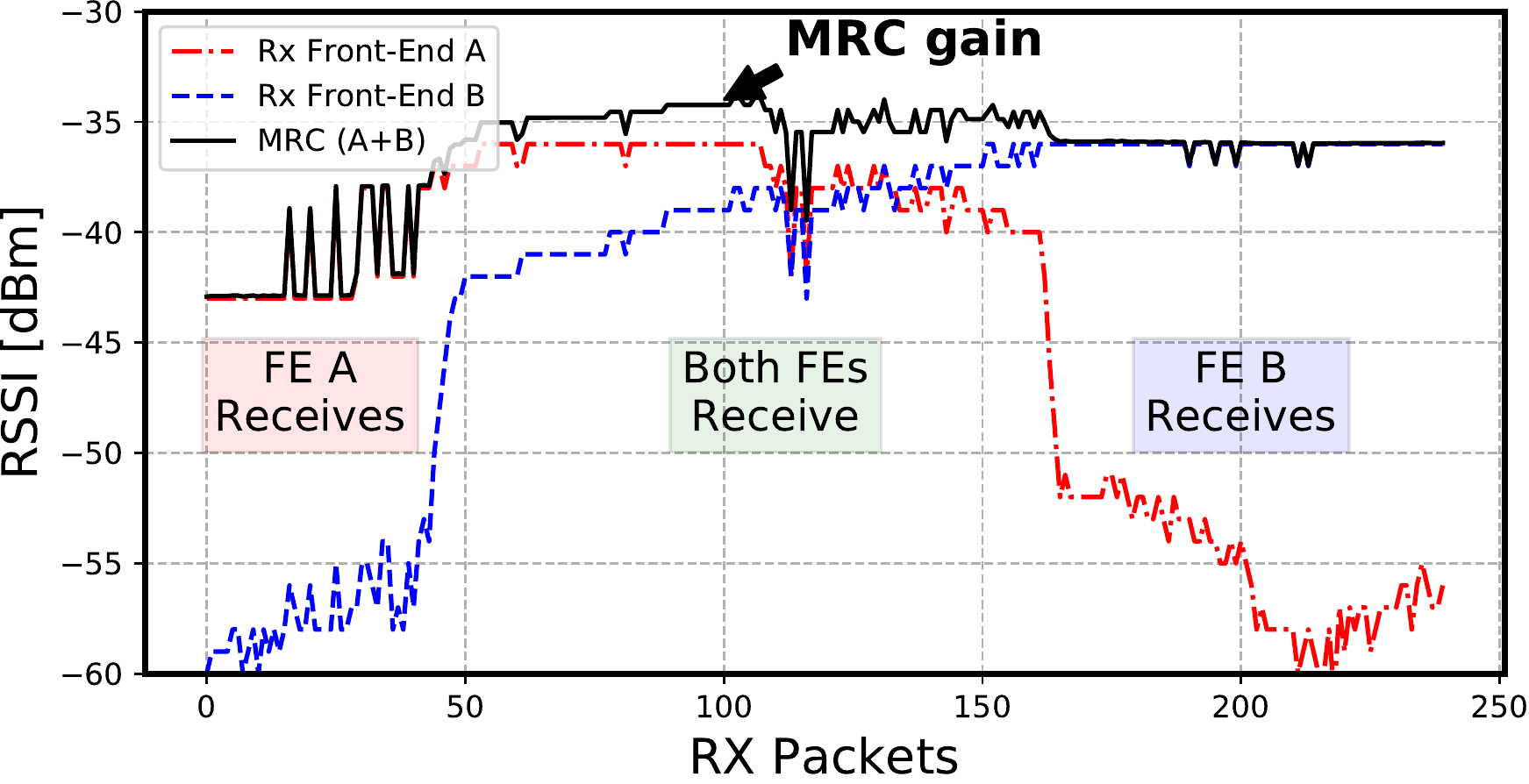}
  \end{minipage}\hfill
  \caption{Distributed antennas for soft-handover.}
  \label{fig:rx_handover}
  \vspace{-5pt}
\end{figure}

The objective of the experiment is to show that receive diversity from 802.11 can be used to implement and support seamless mobility in VLC systems.
We use the same hardware setup as in the previous experiment.
Moreover, we assume that our two RX VLC front-ends are well separated in space, but their signals can be combined using MRC forming some kind of distributed antenna array.
At the beginning of our experiment, we direct the VLC TX front-end to one of the RX front-ends and slowly change its angle-of-view (AoV) (i.e. rotate) the TX front-end towards the second RX front-end (Fig.~\ref{fig:rx_handover}, left).

\noindent\textbf{Results: }
From Fig.~\ref{fig:rx_handover} we can clearly see the gain from spatial receive diversity, i.e. although the AoV of the transmitter changed the combined receive power stayed more or less the same which is in contrast to the two cases where only a single path (i.e. SISO) is used. 
Note that depending on the distance between the distributed VLC RX front-ends and their field-of-view (FOV), the VLC operation may be uninterrupted during such handover, i.e. no network outage.

\subsection{MIMO Performance (Spatial Multiplexing)}
The purpose of this experiment is to evaluate the performance of the spatial multiplexing using our prototype in 2$\times$2 MIMO configuration.
Here, in comparison to the previous setup, we add one more TX VLC front-end and connect it to the second antenna port of the WoV transmitter. 
Note, that the achieved 2$\times$2 MIMO is only uni-directional, i.e. open-loop MIMO without any feedback from the receiver.
During the experiment, we put the RX VLC front-ends in one of the areas:
\begin{itemize}
    \item \textbf{Area 1}: only signal from TX front-end A is received 
    \item \textbf{Area 2}: both signals from TX front-ends are received
    \item \textbf{Area 3}: only signal from TX front-end B is received
\end{itemize}

\noindent For each location, we measure frame success rate (FSR) for different MCSs that utilize two spatial streams. %, i.e. MCS index 8 to 15 in 802.11n.
Moreover, we collect channel state information (CSI) as reported by our Intel WiFi 5300 NIC. 
As it is clear that two spatial streams cannot be received if both RX front-ends are placed in Area 1 or Area 3, we do not analyze those configurations.

\begin{figure}[ht]
  \vspace{-5pt}
  \centering
  \hfill
  \begin{minipage}[ht!]{0.3\linewidth}
    \includegraphics[width=1.0\linewidth]{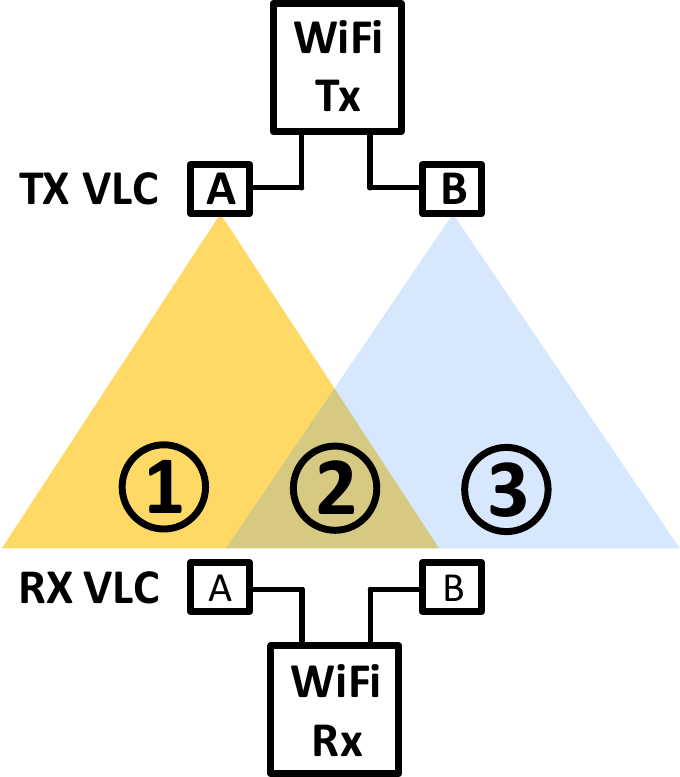}
    %\caption{Experiment Setup.}
  \end{minipage}\hfill
  \begin{minipage}[ht!]{0.7\linewidth}
    \includegraphics[width=1.0\linewidth]{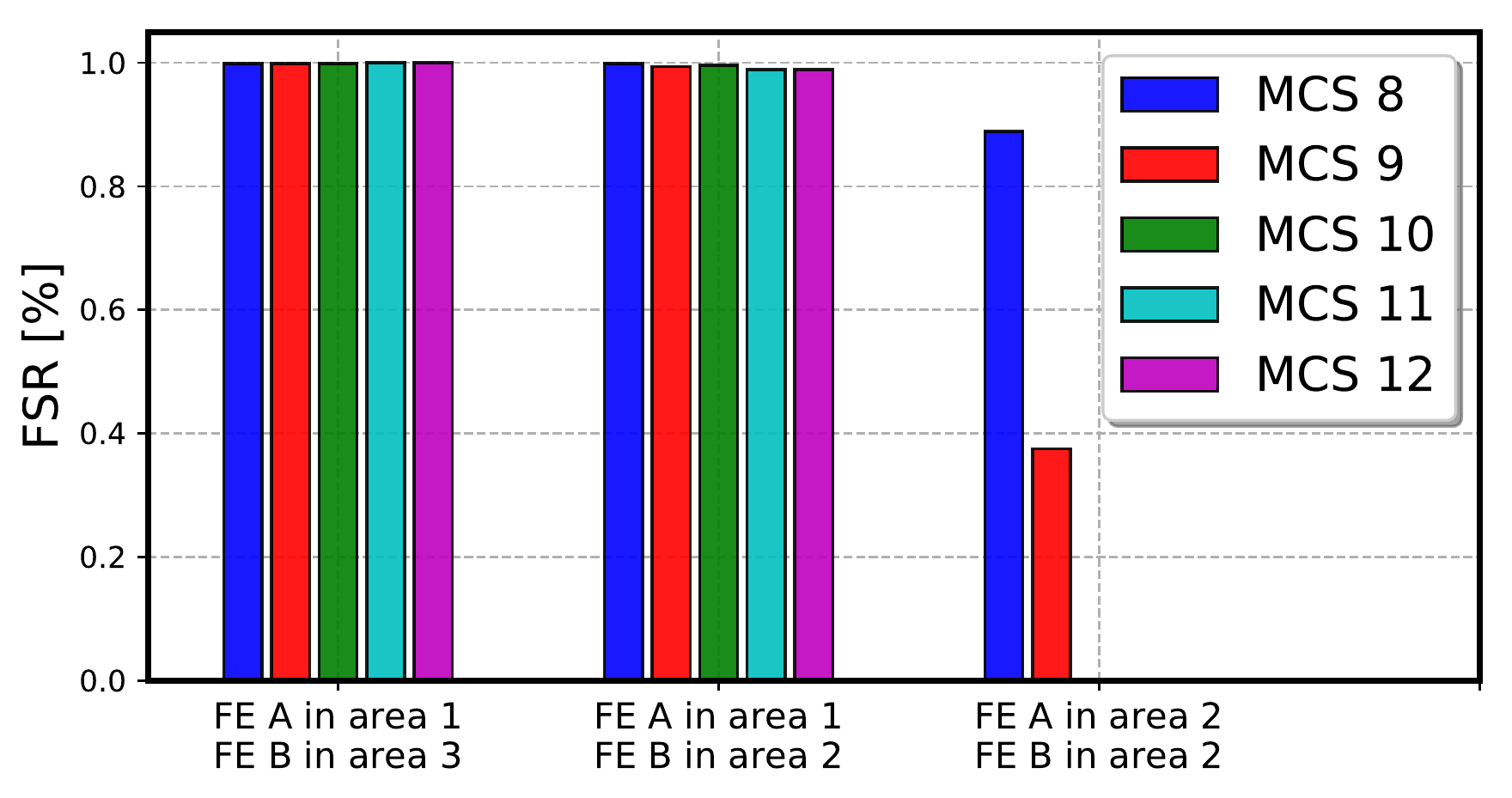}
    %\caption{Frame Success Rate for different VLC front-end configuration.}
  \end{minipage}\hfill
  \caption{Experiment Setup and Frame Success Rate of MIMO transmissions for different VLC front-ends positions.}
  \label{fig:mimo_mcs_fsr}
  \vspace{-5pt}
\end{figure}

\noindent\textbf{Results: }
In Fig.~\ref{fig:mimo_mcs_fsr}, we show obtained results for 802.11n MCSs between 8 (BPSK 1/2) and 12 (16-QAM 3/4).
We can see that in the case of two independent optical wireless links (i.e. when one RX front-end is located in area 1 and the other in area 3) the FSR equals 1.0 for all considered MCSs.
What is even more interesting, if we put one front-end in area 2 and the other one either in area 1 or 3, the FSR is also close to the value of 1.0.
Note that the former RX front-end receives a linear combination of signals coming from TX A and B, while the latter one provides signal only from one TX. 
Therefore, the WiFi receiver is able to solve the MIMO equations and successfully decode the two streams.
Finally, if we put two RX front-ends in area 2, they both receive linear combinations of both streams. 
Therefore, the MIMO equations are not independent and the WiFi receiver cannot solve them. 
As a result, FSR drops to a value of zero due to inter-stream interference. Note that probably due to small difference in receive power of both streams, the MCS 8 (i.e. BPSK 1/2) still provides FSR of more than 0.8.

\begin{figure*}[ht]
  \vspace{0pt}
  \begin{minipage}[b]{0.25\linewidth}
    \caption*{\footnotesize FE A in area 1 \& FE B in area 3}
    \vspace{-5pt}
	\includegraphics[width=\linewidth]{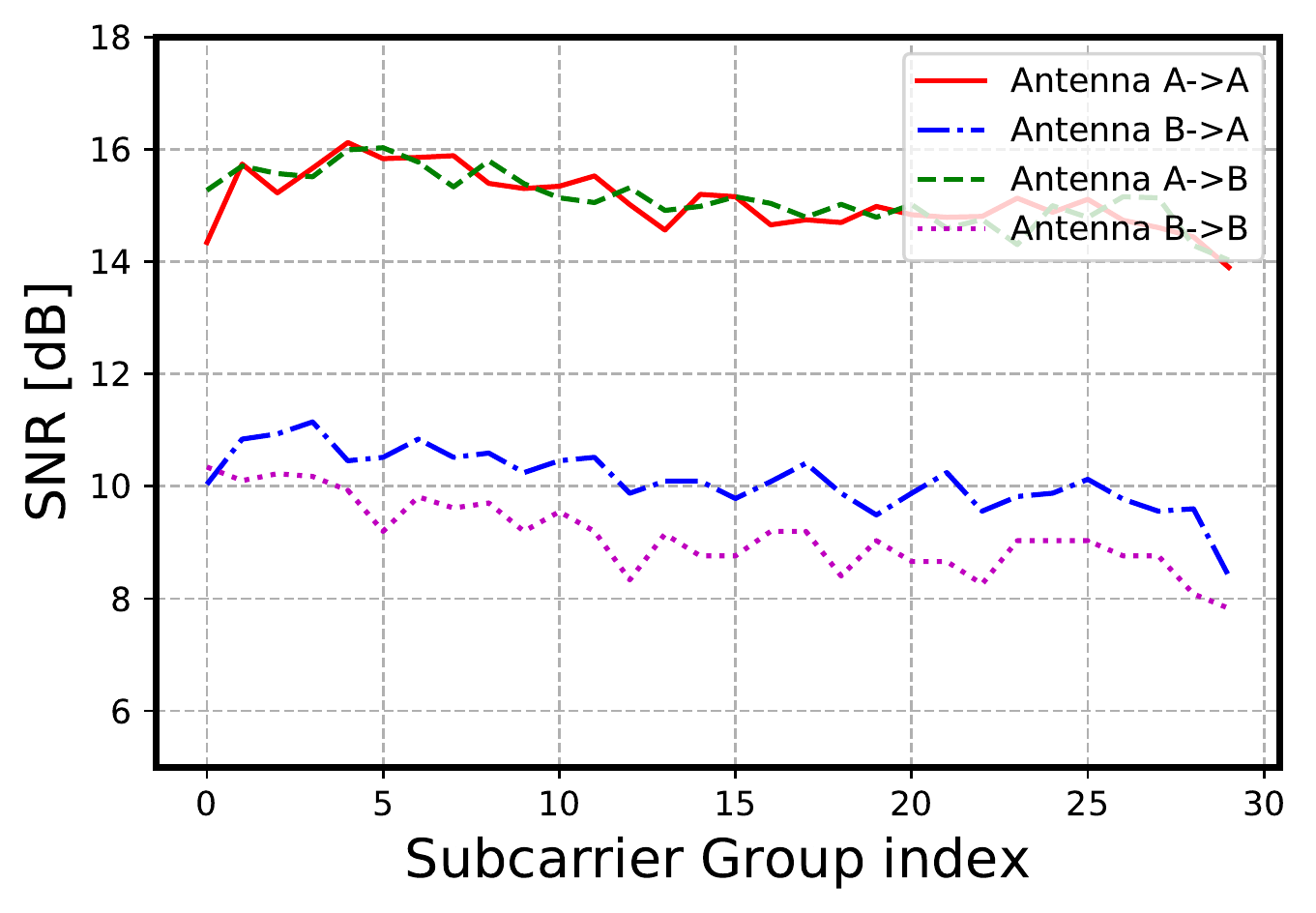}
    \label{fig:constDiagram_A}
  \end{minipage}\hfill
  \begin{minipage}[b]{0.25\linewidth}
    \caption*{\footnotesize FE A in area 2 \& FE B in area 3}
    \vspace{-5pt}
	\includegraphics[width=\linewidth]{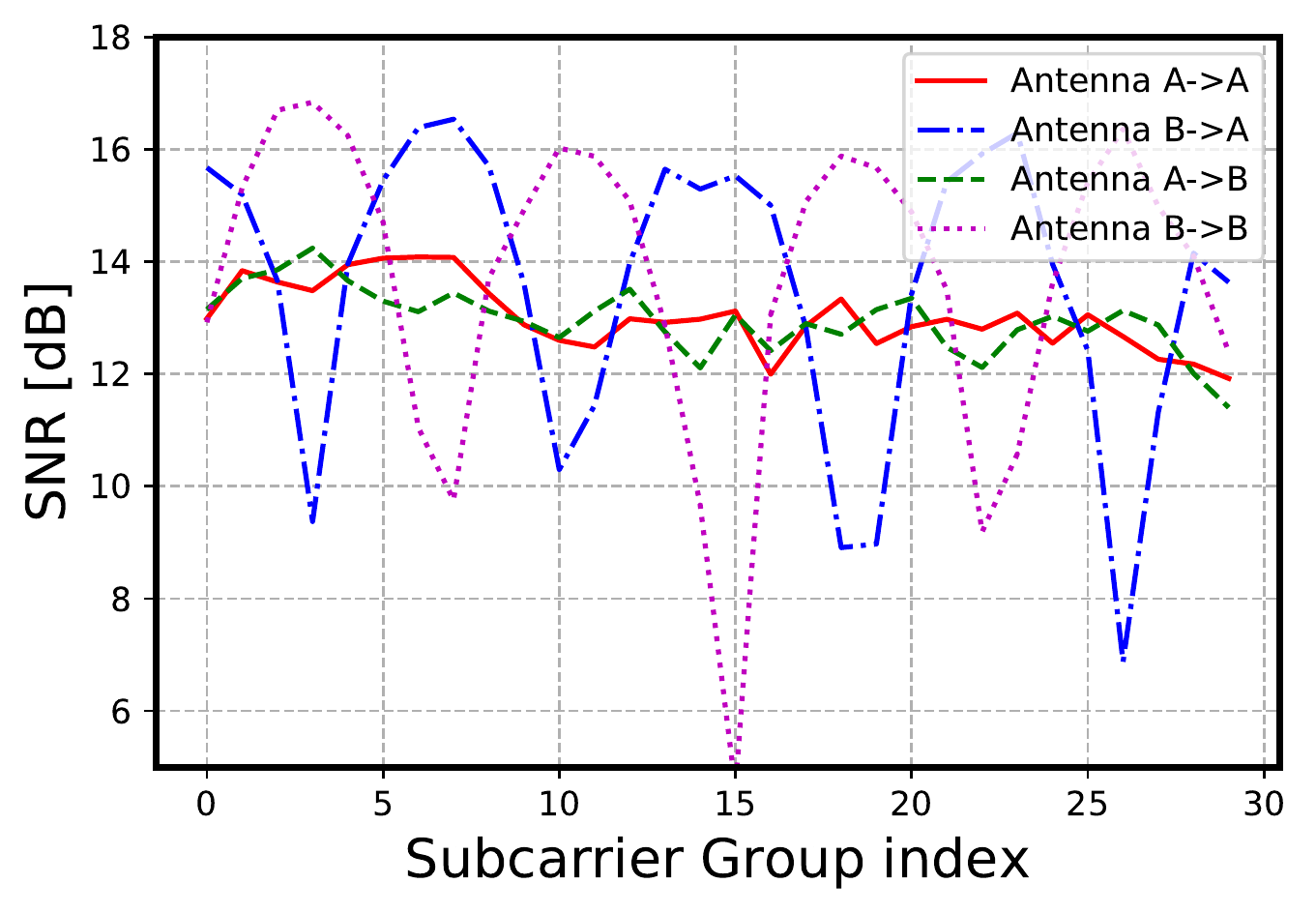}
    \label{fig:constDiagram_B}
  \end{minipage}\hfill
  \begin{minipage}[b]{0.25\linewidth}
    \caption*{\footnotesize FE A in area 1 \& FE B in area 2}
    \vspace{-5pt}
	\includegraphics[width=\linewidth]{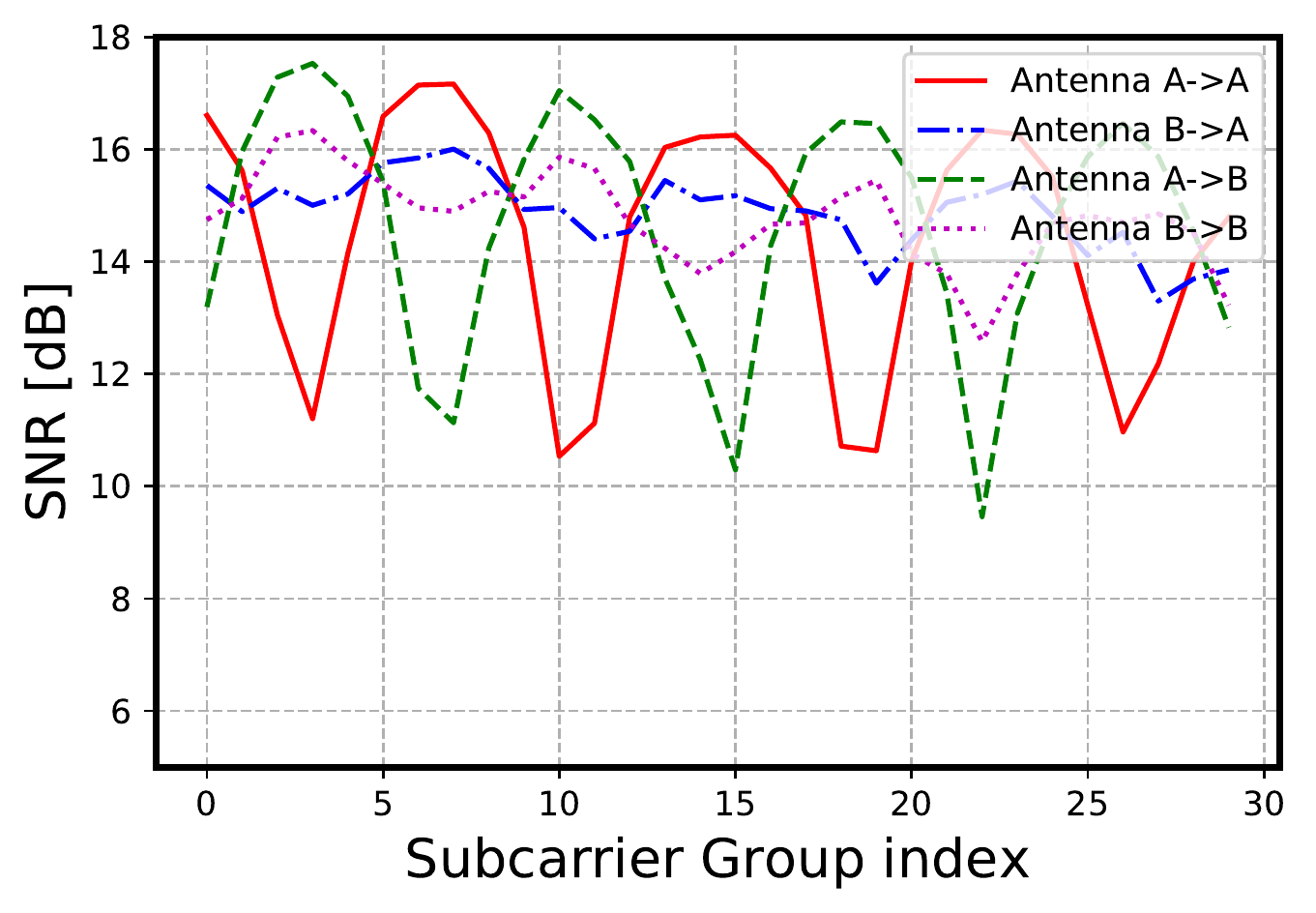}
    \label{fig:constDiagram_D}
  \end{minipage}\hfill
  \begin{minipage}[b]{0.25\linewidth}
    \caption*{\footnotesize FE A in area 2 \& FE B in area 2}
    \vspace{-5pt}
	\includegraphics[width=\linewidth]{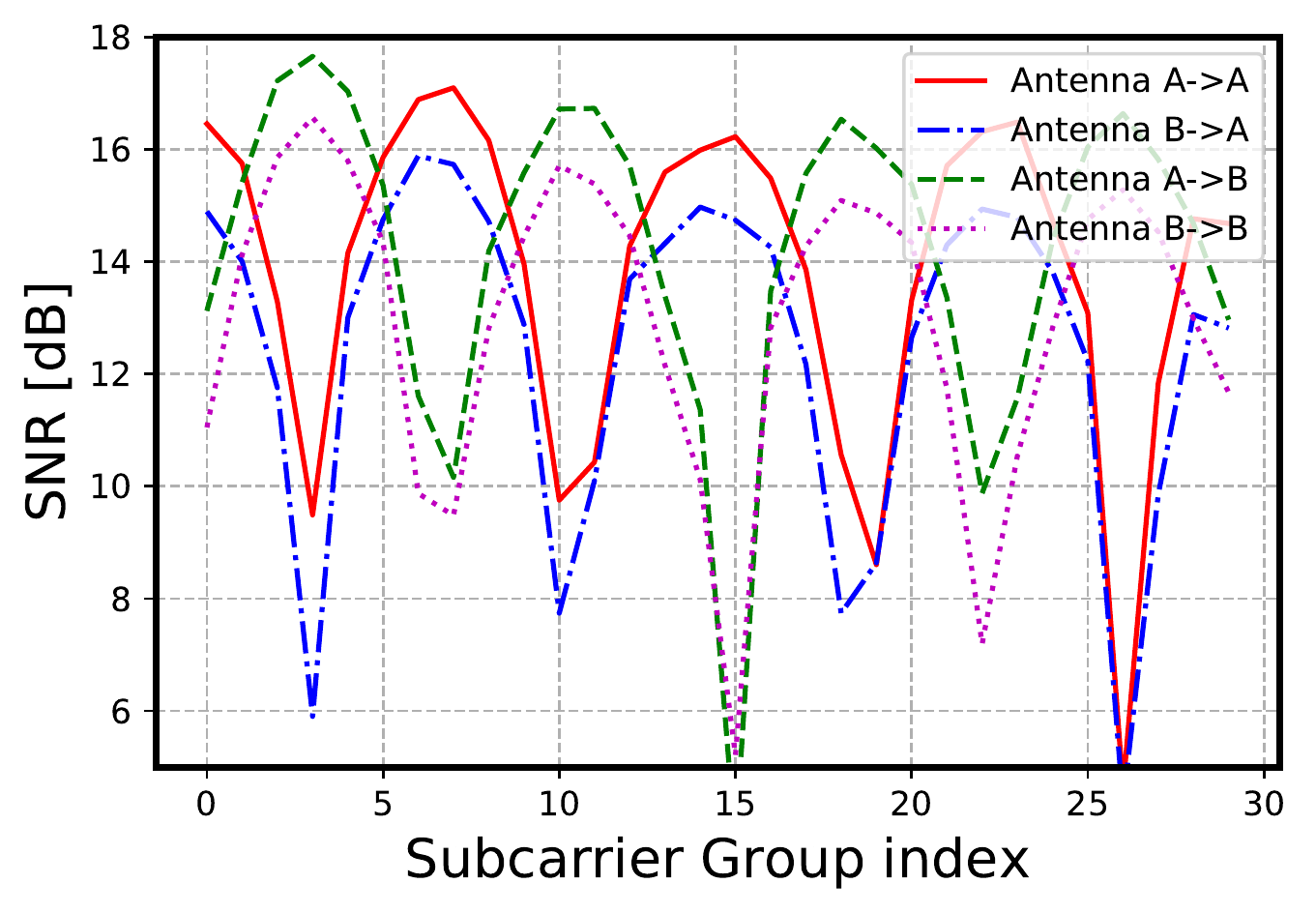}
    \label{fig:constDiagram_E}
  \end{minipage}\hfill
  \vspace{-15pt}
  \caption{Channel State Information (CSI) for different VLC front-end receiver locations.}
  \label{fig:mimo_csis}
  \vspace{-10pt}
\end{figure*}

The CSI as reported by the WiFi NIC on the receiver side is shown in Fig.~\ref{fig:mimo_csis}. Note that frames were transmitted with MCS 0 (i.e. single stream over two TX front-ends).
We can see that independent SISO VLC channels (i.e. area 1 and 2) are frequency flat (the 3\,dB noise in the figures is caused by the low quality, i.e. 6-bits, quantization). 
However, when an RX front-end receives signals from two TX front-ends in MISO and/or MIMO configurations (i.e. in area 2), the perceived channel becomes frequency-selective. 
This is an expected effect, i.e. the usage of multiple TX is comparable to multi-path propagation~\cite{tse2005fundamentals}.

%
%%%
\section{Discussion}\label{sec:discussion}
With our MIMO-WoV testbed, we have shown that the advanced signal processing techniques already available in today's WiFi devices can be successfully reused in VLC technology.
However, our prototype not only inherits all the limitations of the WoV platform~\cite{gawowicz2020wov} (e.g., we can utilize only WiFi signals in 2.4\,GHz band, where the maximal channel bandwidth is just 40\,MHz as compared to up-to 160\,MHz available in 5\,GHz spectrum), but also the usage of antenna diversity techniques reveals additional shortcomings of the current version.
Namely, a single RF chain of WiFi NIC serves both TX and RX directions, but we restrict it only to one of those two by properly changing the configuration of the NIC. 
This way, it was very easy and inexpensive to turn a WiFi NIC into a WoV VLC transceiver. However, we are artificially restricting MIMO capabilities of the WiFi device, e.g. out of a 2x2 MIMO WiFi transceiver, we make a 1x1 SISO VLC transceiver (i.e. one bi-directional link).
In this work, we present and evaluate up to 2x2 VLC MIMO transmissions, but only in uni-directional link configuration, i.e. one WoV node with two TX chains and one WoV node with two RX chains.
A feasible solution, in order to avoid MIMO dimension reduction, is to employ an RF circulator (or two RF isolators) in our signal up-/down-conversion setup and connect both VLC TX and RX front-ends to a single TX/RX port of WiFi NIC. Those components prevent a signal from PD to be leaked and transmitted by LED and a signal from WiFi NIC to disturb the operation of (or possibly destroy) the PD.

Furthermore, in order to build large MIMO VLC systems using the WoV platform, the cost of a single signal adapting module has to be minimized. The current cost is around 130\$~\cite{gawowicz2020wov}, for a single TX/RX VLC chain (without the RF circulator). We believe that the design and development of a printed circuit board (PCB) integrating all of the used components will cut the cost by a factor of 2-3 times.

%
%%%
\section{Related Work}

Recent work confirms that there is a huge interest in applying MIMO techniques into optical wireless communication~\cite{AlAhmadi2018MultiUserVL,Pathak2015VisibleLC}. 
However, due to a lack of evaluation platforms and/or its high complexity and cost, a lot of research is performed by means of simulation analysis only. 
For example in~\cite{7794946}, Ying, et al. provides a design of a MIMO VLC transceiver and evaluates its performance by numerical analysis.
In~\cite{6916680} a simulation study of the performance of repetition code and spatial multiplexing applied to VLC in indoor scenarios with non-direct line-of-sight (LOS) was presented. 
In~\cite{7109107}, the authors propose a new approach for designing the layout of photo-detector (PD) arrays in the MIMO-VLC receiver. 
Specifically, by varying the orientation angles of the PDs, they achieve highly uncorrelated optical channels and hence multiplexing gain by angle diversity. 
The new PD arrays were evaluated by means of simulation and in a small custom-made testbed.
As our MIMO-WoV platform is based only on inexpensive COTS devices, the above-mentioned approaches can be verified in an experimental testbed. 
Therefore, we believe that our work will encourage researchers to test their VLC ideas in real-world experiments.
We are aware that our platform is restricted only to WiFi capabilities, however, the set of implemented signal processing schemes already covers most of the popular ones including low-density parity-check coding (LDPC), Space-Time Block Coding (STBC) or beamforming techniques like Multi-user MIMO.
Therefore, researchers do not have to repeat the tedious task of reimplementing those algorithms in order to perform measurements, as it was done for example in~\cite{8742557}.

Furthermore, a number of custom-built MIMO-VLC platforms were presented recently.
For example, using proprietary MIMO transceiver Berenguer et al.~\cite{berenguer2017optical} performed measurements in the manufacturing environment. Their results proved that antenna diversity over MIMO channels is indispensable for reliable optical wireless communications as it copes with LOS-blockage that is frequent in industrial environments.
In~\cite{7936916} authors demonstrated an SDR-based 2$\times$2 MIMO VLC system employing spatial multiplexing and diversity.
While, Werfli et al.\cite{8265090} investigated the performance of a 4$\times$4 MIMO VLC transceiver reaching a maximum data rate of 249\,Mbps. Note that with our 2$\times$2 MIMO-WoV system, we achieve up to 300\,Mbps using 40\,MHz channels.
DenseVLC~\cite{DenseVLC} is a cell-free massive (CFM) MIMO system enabled by densely distributed LEDs. It consists of low-cost and densely distributed OpenVLC~\cite{Galisteo2} VLC transceivers and can enable adaptive and power-efficient CFM-MIMO beamspots to serve multiple receivers simultaneously. However, OpenVLC uses simple \textit{on-off keying} modulation providing a maximal data rate of up to 400\,kbps. 
Finally, a similar WiFi-based VLC platform, however, restricted only to the evaluation of SISO links, was presented in~\cite{amjad2019ieee}.

%
%%%
\section{Conclusions}

We present MIMO-WoV, a first VLC platform that reuses advanced signal processing techniques (i.e. spatial diversity and spatial multiplexing) available in modern COTS RF WiFi devices for enhancing the transmissions over optical wireless links. 
Specifically, we have shown that the VLC technology, which is still in its early stages, can be brought to maturity in no time by simply exploiting knowledge and years of research and development spent on WiFi.
Our evaluation confirms that the spatial receive diversity and MIMO techniques are beneficial for VLC, as they can effectively mitigate signal blockage issues and boost the capacity of the VLC system, respectively.
The platform is inexpensive and easy to extend, hence, we believe it will encourage and speed up further research and development in the area of VLC research.

As future work, we plan to develop and evaluate MIMO-WoV transceiver with support of higher-order MIMO (e.g. 4$\times$4) as well as analyze the performance of advanced multi-user beamforming and transmission techniques like Multi-user MIMO and OFDMA from 802.11ac/ax standards over VLC.
Moreover, spatial diversity can be obtained not only at the physical layer but also at the MAC layer~\cite{miu2005improving}.
It was shown that the technique termed as MAC Diversity is very efficient in diminishing the adverse effects of Shadowing and interference in RF systems like 802.11~\cite{zubow2012mac}.
Here we plan to use our testbed to quantify the gain from MAC diversity for both SISO systems as well as MIMO, i.e. in addition to PHY diversity.

\section*{Acknowledgement}
We are grateful to Fraunhofer HHI for providing us the VLC front-ends.
This work was supported by the German BMBF under grant agreement No. 16KIS0985 (OTB-5G+ project).

\bibliographystyle{IEEEtran}
\footnotesize{\bibliography{biblio,IEEEabrv}}

\begin{thebibliography}{10}
\providecommand{\url}[1]{#1}
\csname url@samestyle\endcsname
\providecommand{\newblock}{\relax}
\providecommand{\bibinfo}[2]{#2}
\providecommand{\BIBentrySTDinterwordspacing}{\spaceskip=0pt\relax}
\providecommand{\BIBentryALTinterwordstretchfactor}{4}
\providecommand{\BIBentryALTinterwordspacing}{\spaceskip=\fontdimen2\font plus
\BIBentryALTinterwordstretchfactor\fontdimen3\font minus
  \fontdimen4\font\relax}
\providecommand{\BIBforeignlanguage}[2]{{%
\expandafter\ifx\csname l@#1\endcsname\relax
\typeout{** WARNING: IEEEtran.bst: No hyphenation pattern has been}%
\typeout{** loaded for the language `#1'. Using the pattern for}%
\typeout{** the default language instead.}%
\else
\language=\csname l@#1\endcsname
\fi
#2}}
\providecommand{\BIBdecl}{\relax}
\BIBdecl

\bibitem{jovicic2013visible}
A.~Jovicic, J.~Li, and T.~Richardson, ``Visible light communication:
  opportunities, challenges and the path to market,'' \emph{IEEE Communications
  Magazine}, vol.~51, no.~12, pp. 26--32, 2013.

\bibitem{ghassemlooy2019optical}
Z.~Ghassemlooy, W.~Popoola, and S.~Rajbhandari, \emph{Optical wireless
  communications: system and channel modelling with
  Matlab{\textregistered}}.\hskip 1em plus 0.5em minus 0.4em\relax CRC press,
  2019.

\bibitem{tse2005fundamentals}
D.~Tse and P.~Viswanath, \emph{Fundamentals of wireless communication}.\hskip
  1em plus 0.5em minus 0.4em\relax Cambridge university press, 2005.

\bibitem{halperin2009two}
D.~Halperin, W.~Hu, A.~Sheth, and D.~Wetherall, ``Two antennas are better than
  one: A measurement study of 802.11 n,'' \emph{Unpublished technical report,
  University of Washington}, 2009.

\bibitem{berenguer2017optical}
P.~W. Berenguer, D.~Schulz, J.~Hilt, P.~Hellwig, G.~Kleinpeter, J.~K. Fischer,
  and V.~Jungnickel, ``Optical wireless mimo experiments in an industrial
  environment,'' \emph{IEEE Journal on Selected Areas in Com.}, 2017.

\bibitem{AlAhmadi2018MultiUserVL}
S.~Al-Ahmadi, O.~Maraqa, M.~Uysal, and S.~M. Sait, ``{Multi-User Visible Light
  Communications: State-of-the-Art and Future Directions},'' \emph{IEEE
  Access}, 2018.

\bibitem{gawowicz2020wov}
\BIBentryALTinterwordspacing
P.~Gawłowicz, E.~A. Jarchlo, and A.~Zubow, ``{WoV: WiFi-based VLC testbed},''
  \emph{CoRR}, 2020. [Online]. Available:
  \url{https://arxiv.org/abs/2001.08489}
\BIBentrySTDinterwordspacing

\bibitem{halperin2011predictable}
D.~Halperin, W.~Hu, A.~Sheth, and D.~Wetherall, ``Predictable 802.11 packet
  delivery from wireless channel measurements,'' \emph{ACM SIGCOMM}, 2011.

\bibitem{Pathak2015VisibleLC}
P.~H. Pathak, X.~Feng, P.~Hu, and P.~Mohapatra, ``{Visible Light Communication,
  Networking, and Sensing: A Survey, Potential and Challenges},'' \emph{{IEEE
  Communications Surveys \& Tutorials}}, 2015.

\bibitem{7794946}
Z.~{Wang}, C.~{Guo}, Y.~{Yang}, and Q.~{Li}, ``{Antenna selection based dimming
  scheme for indoor MIMO visible light communication systems utilizing multiple
  lamps},'' in \emph{IEEE PIMRC}, 2016.

\bibitem{6916680}
{Ngoc-Anh Tran}, D.~A. {Luong}, T.~C. {Thang}, and A.~T. {Pham}, ``{Performance
  analysis of indoor MIMO visible light communication systems},'' in \emph{IEEE
  ICCE}, 2014.

\bibitem{7109107}
A.~{Nuwanpriya}, S.~{Ho}, and C.~S. {Chen}, ``{Indoor MIMO Visible Light
  Communications: Novel Angle Diversity Receivers for Mobile Users},''
  \emph{{IEEE Journal on Selected Areas in Com.}}, 2015.

\bibitem{8742557}
A.~{Khalid}, H.~M. {Asif}, S.~{Mumtaz}, S.~{Al Otaibi}, and K.~{Konstantin},
  ``{Design of MIMO-Visible Light Communication Transceiver Using Maximum Rank
  Distance Codes},'' \emph{IEEE Access}, 2019.

\bibitem{7936916}
P.~{Deng} and M.~{Kavehrad}, ``{Software defined adaptive MIMO visible light
  communications after an obstruction},'' in \emph{Optical Fiber Communications
  Conference}, 2017.

\bibitem{8265090}
K.~{Werfli}, P.~{Chvojka}, Z.~{Ghassemlooy}, N.~B. {Hassan}, S.~{Zvanovec},
  A.~{Burton}, P.~A. {Haigh}, and M.~R. {Bhatnagar}, ``{Experimental
  Demonstration of High-Speed 4 $\times$ 4 Imaging Multi-CAP MIMO Visible Light
  Communications},'' \emph{Journal of Lightwave Technology}, 2018.

\bibitem{DenseVLC}
J.~Beysens, A.~Galisteo, Q.~Wang, D.~Juara, D.~Giustiniano, and S.~Pollin,
  ``{DenseVLC: A Cell-Free Massive MIMO System with Distributed LEDs},'' in
  \emph{ACM CoNEXT}, 2018.

\bibitem{Galisteo2}
A.~{Galisteo}, D.~{Juara}, and D.~{Giustiniano}, ``Research in visible light
  communication systems with openvlc1.3,'' in \emph{IEEE WF-IoT}, 2019.

\bibitem{amjad2019ieee}
M.~S. Amjad, C.~Tebruegge, A.~Memedi, S.~Kruse, C.~Kress, C.~Scheytt, and
  F.~Dressler, ``{An IEEE 802.11 Compliant SDR-based System for Vehicular
  Visible Light Communications},'' 2019.

\bibitem{miu2005improving}
A.~Miu, H.~Balakrishnan, and C.~E. Koksal, ``Improving loss resilience with
  multi-radio diversity in wireless networks,'' in \emph{ACM MobiCom}, 2005.

\bibitem{zubow2012mac}
A.~Zubow, R.~Sombrutzki, and M.~Scheidgen, ``Mac diversity in ieee 802.11 n
  mimo networks,'' in \emph{IFIP Wireless Days}.\hskip 1em plus 0.5em minus
  0.4em\relax IEEE, 2012.

\end{thebibliography}
\end{document}